\documentclass[reprint,amsmath,amssymb,aps,prl,floatfix]{revtex4-2}
\pdfoutput=1

\usepackage{graphicx}%
\usepackage{amsmath}
\usepackage{hyperref}
\usepackage{cleveref}
\newcommand{\smcite}{}

\begin{document}
\preprint{APS/123-QED}

\title{Reproducible distributions of energy-relaxation times in superconducting qubits from broadband SWAP spectroscopy}

\author{Joaquin Minguzzi, John Clai Owens, Maximilien F. Debbas, Ana Valdés-Curiel, Ming-Han Chou, Haifeng Qiao, Shahriar Aghaeimeibodi, Oskar Painter, Matthew H. Matheny, Loren J. Swenson}
\affiliation{AWS Center for Quantum Computing, Pasadena, CA 91106, USA}

\date{\today}

\begin{abstract}
The distribution of the qubit energy-relaxation time ($T_1$) in a superconducting quantum processor affects its logical error rates.
However, reports on qubit fabrication quality focus on the median or best-case $T_1$ from measurements that are not reproducible due to resonant coupling to two-level system (TLS) defects whose frequencies fluctuate in time.
Here, we sweep the frequency $f_{01}$ of a flux-tunable qubit over several GHz with SWAP spectroscopy, quoting the quality factor $Q = 2\pi f_{01} T_1$ to make $T_1$ comparable across frequencies.
We show that the cumulative distribution of $Q$ pooled over the sweep is reproducible across four thermal cycles spanning 326~days, up to the reshuffling of strongly coupled TLSs.
In contrast to repeated measurements at a single frequency, the medians of the broadband $Q$ distributions spread by an order of magnitude less over 3~days.
The distance between the cumulative distributions of repeats comes close to its sampling floor only for the broadband sweeps.
The broadband $Q$ distribution further resolves small coherence differences between identically designed qubits on two dies of the same wafer.
These results establish the broadband $Q$ distribution as a reproducible metric for qubit fabrication quality.
\end{abstract}

\maketitle

\begin{figure}[!t]
\centering
\includegraphics[width=0.95\columnwidth]{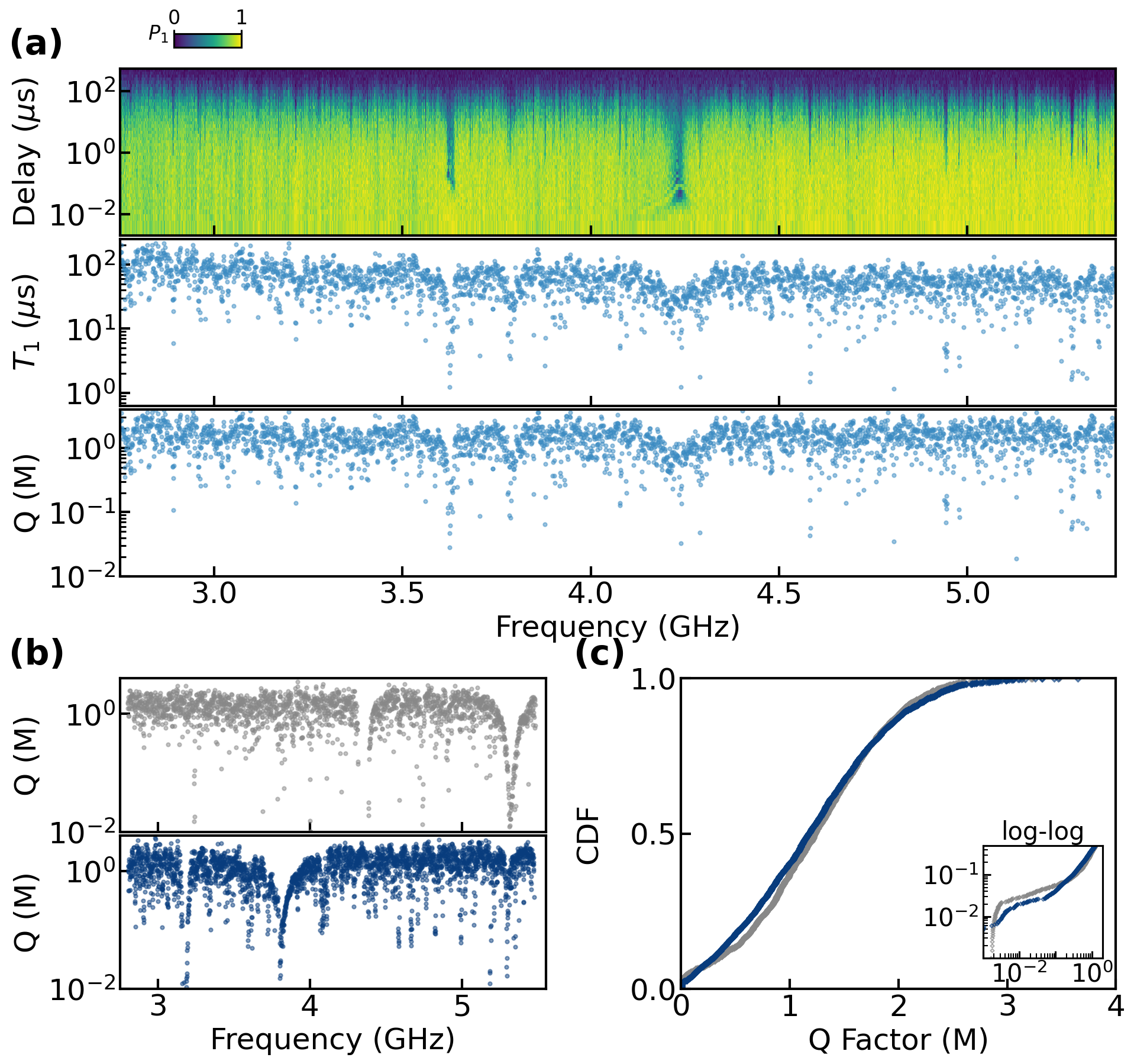}
\caption{\protect\input{figure1_caption.tex}}
\label{fig:fig1}
\end{figure}

Improvements in the coherence times of superconducting qubits~\cite{Siddiqi2021} must be reproducible in time and uniform across fabrication scales, from the qubits of a processor to dies across a wafer~\cite{VanDamme2024,Mayer2025,Mohseni2024}, and so call for efficient cryogenic characterization.
The qubit energy-relaxation time ($T_1$) is typically characterized through summary statistics of inversion recovery experiments repeated over hours or days.
However, differences between fabrication processes can be masked by the large temporal variability of $T_1$~\cite{Burnett2019}.
This variability is due to two-level system (TLS) defects~\cite{Leggett2013,Mueller2019,Yu2021}, whose spectral dynamics are driven by their mutual interactions~\cite{FaoroIoffe2012,Lisenfeld2015,Klimov2018} and which reshuffle across thermal cycles~\cite{Zanuz2025}.
Sweeping the qubit transition frequency $f_{01}$, as in AC-Stark shift spectroscopy~\cite{Carroll2022,Thorbeck2023} and SWAP spectroscopy~\cite{Barends2013,Klimov2018}, samples many TLS defects in a single experiment.
Coherent swapping with strongly coupled TLSs produces broad loss features~\cite{Martinis2005}, while finer ones arise from incoherent decay into a sparse population of weakly coupled TLSs~\cite{Barends2013}; see Fig.~\ref{fig:fig1}a.
Such sweeps, repeated over many hours, then characterize the qubit energy relaxation through a single frequency-averaged $T_1$~\cite{Carroll2022,Weeden2025}.
Similarly, modulating the frequencies of surface TLS defects with an applied electric field enhances the measured $T_1$ or stabilizes it in time~\cite{Lisenfeld2019,Lisenfeld2023,Dane2026}.

These methods, however, require long measurement times per qubit, and have focused on the frequency-averaged $T_1$, drawn from distributions that are themselves not stationary, that is, not reproducible in time.
In this work, we show that the $T_1$ distribution from a single SWAP spectroscopy sweep, broad in bandwidth and fine in frequency step, approaches stationarity.
The bandwidth is broad enough to cover the few strongly coupled TLSs, and the frequency step fine enough to resolve the weakly coupled ones.
Over the swept bandwidth, we quote the quality factor $Q = 2\pi f_{01} T_1$ in place of $T_1$, and we keep every point of the sweep rather than a summary statistic.
We find that the cumulative distribution function (CDF) of those $Q$ values is repeatable within one thermal cycle and reproducible across four thermal cycles spanning 326~days, up to the reshuffling of strongly coupled TLSs.
We compare this with a narrower sweep equivalent to AC-Stark shift spectroscopy and with repeated inversion recovery for the same measurement time, and neither reaches that repeatability.
With the broadband $Q$ CDF we then resolve small differences in the coherence times of identically designed qubits between two dies of the same wafer.

We used the fast-flux-assisted (FFA) SWAP spectroscopy technique introduced in Ref.~\cite{Minguzzi2026} to collect broadband $T_1$ spectra (Fig.~\ref{fig:fig1}a).
We operated aluminum-on-silicon transmon qubits with flux-tunable $f_{01}$ over a bandwidth of approximately 2.7~GHz extending down from the flux-insensitive point.
In the FFA SWAP pulse sequence, a DC bias tunes $f_{01}$ along the tuning curve and fast-flux pulses bring it to the flux-insensitive point for state preparation and readout.
Switching the fast flux off during energy relaxation reduces flux noise to the level set by the DC bias alone, giving the stability needed to probe the qubit $T_1$ in fine frequency steps (see Sec.~\ref{sec:supp_ffa} of the Supplemental Material\smcite).
Fitting an exponential decay to the relaxation of the qubit excited-state population $P_1$, logarithmically sampled at 48 delays from 4~ns to 500~\textmu{}s, yields the $T_1$ spectrum at each frequency (Fig.~\ref{fig:fig1}a, middle), with dropouts as low as $\sim$1~\textmu{}s against a baseline of $\sim$100~\textmu{}s.
The baseline falls as $1/f_{01}$, as predicted by the standard tunneling model~\cite{Phillips1987} for a bath of TLS defects coupled through the electric field in the qubit capacitor gap~\cite{Murray2018,Weeden2025}.
Converting the $T_1$ spectrum to $Q$ removes this frequency dependence (Fig.~\ref{fig:fig1}a, bottom).

Pooling every point of the $Q$ spectrum gives the broadband $Q$ distribution, and sorting the values yields its CDF, which we compare across thermal cycles (TC) to test stationarity.
To illustrate, Fig.~\ref{fig:fig1}b shows the $Q$ spectra of two TCs separated by 211~days and Fig.~\ref{fig:fig1}c their CDFs.
The deepest dropouts sit at different frequencies in the two spectra, so thermal cycling reshuffles the TLS configuration, yet the two CDFs overlap tightly.
The high $Q$ values, in the millions (M), reflect the baseline set by the TLS bath and radiative loss, while the low-$Q$ tail, visible in the log-log inset, captures energy loss from strongly coupled TLSs.
Near a TLS resonance the $P_1$ decay becomes nonexponential, since the qubit and the defect undergo Rabi exchange; fitting that coherent oscillation with an exponential decay returns both a short $T_1$ and a large reduced chi-squared ($\chi^2_r$)~\cite{Minguzzi2026}.
Every $Q$ reported here comes from a fit with $\chi^2_r < 20$ and a $T_1$ uncertainty smaller than $T_1$; retaining only fits with $\chi^2_r < 2$ preferentially removes the low-$Q$ points (see Sec.~\ref{sec:supp_redchi}).

\begin{figure}[!t]
\centering
\includegraphics[width=0.95\columnwidth]{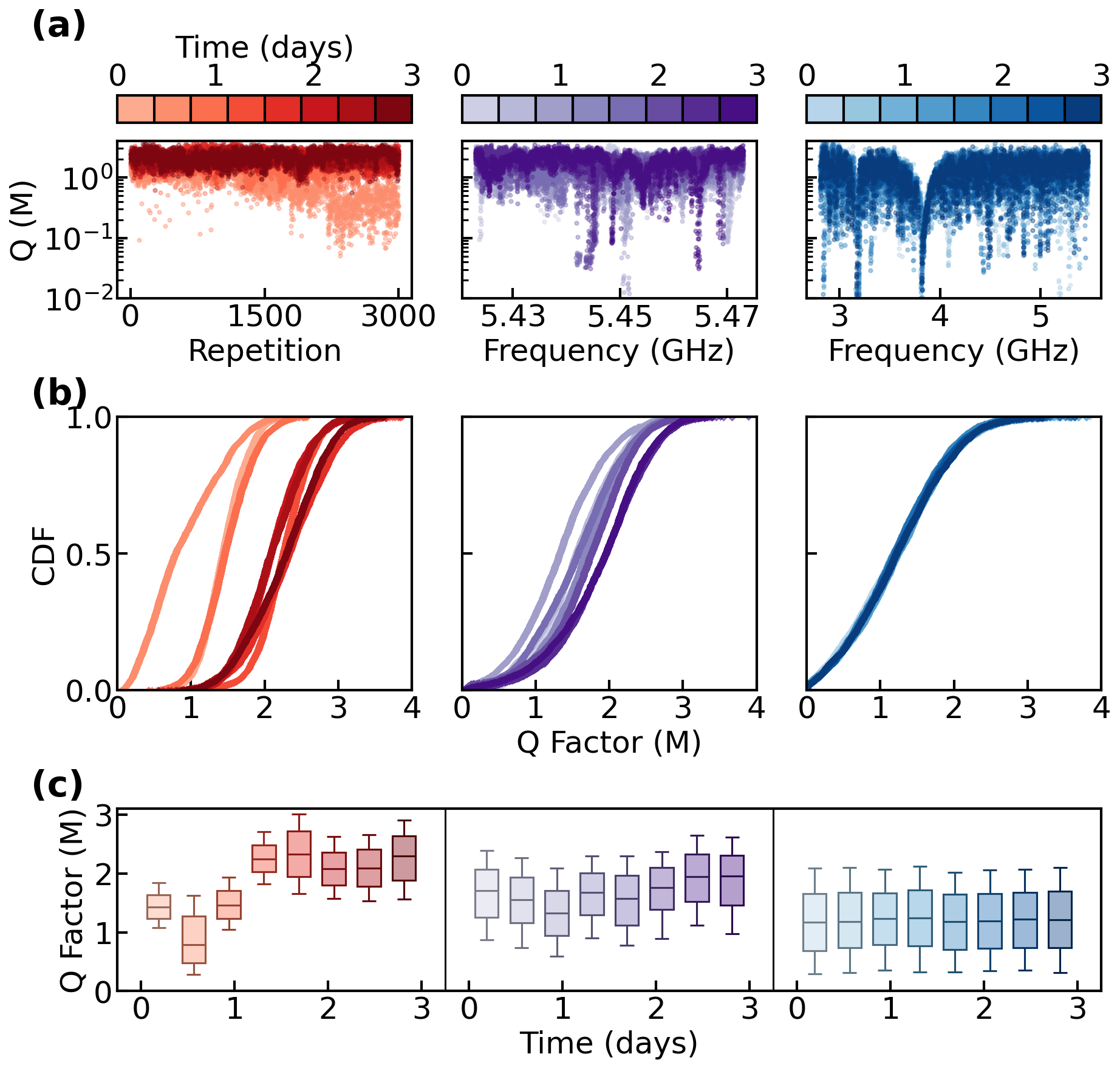}
\caption{\protect\input{figure2_caption.tex}}
\label{fig:fig2}
\end{figure}

To attribute this stationarity to the bandwidth itself, we compared the $Q$ distributions from repeated inversion recovery, from a narrowband sweep equivalent to AC-Stark shift spectroscopy, and from a broadband sweep (Fig.~\ref{fig:fig2}).
We emulated the first two experiments using a single frequency point (zero bandwidth) and a 50~MHz bandwidth from the same FFA SWAP spectroscopy sequence.
Each experiment used 3{,}000 points and the same measurement time, with the two sweeps stepping in frequency from the flux-insensitive point and the single-frequency case repeating 3{,}000 times at that point.
We interleaved these experiments and repeated them eight times at 9~hour intervals over 3~days, so that all three sampled the same TLS spectral dynamics.
Fig.~\ref{fig:fig2}a shows the eight traces of each bandwidth superimposed in one panel per bandwidth, shaded by elapsed time, so that the changes in $Q$ between repeats are visible.
Despite sampling more strongly coupled TLSs, the broadband sweeps show a tighter overlap of their $Q$ CDFs than the narrowband and single-frequency cases (Fig.~\ref{fig:fig2}b).

We extracted summary statistics from each $Q$ distribution to track the time evolution of the broadband sweeps and the narrower-bandwidth alternatives (Fig.~\ref{fig:fig2}c).
The single-frequency distributions show the highest $Q$ but also the widest spread at all percentiles, and the narrowband and broadband distributions are progressively more stable.
For example, the medians of the eight repeats spread by 1.54~M (74\%) for the single-frequency case, 0.63~M (37\%) for the narrowband sweep and 0.07~M (6\%) for the broadband sweep.
Beyond one-sample percentiles, we used the two-sample Kolmogorov-Smirnov (KS) distance, the largest vertical gap between two CDFs.
Among the 28 pairs of repeats of each bandwidth, the median KS distance is 0.57 for the single-frequency case, 0.17 for the narrowband sweep and 0.03 for the broadband sweep, the only one that comes close to the sampling floor of $\sim$0.02, which we measure by permuting the $Q$ values among the repeats~\cite{Feigelson2012}.
Sampling the broadband sweeps with a coarser frequency step raises that floor faster than it raises the median KS distance, so the measurement itself can no longer distinguish the repeats (see Sec.~\ref{sec:supp_bandwidth_ks}).

\begin{figure}[!t]
\centering
\includegraphics[width=0.95\columnwidth]{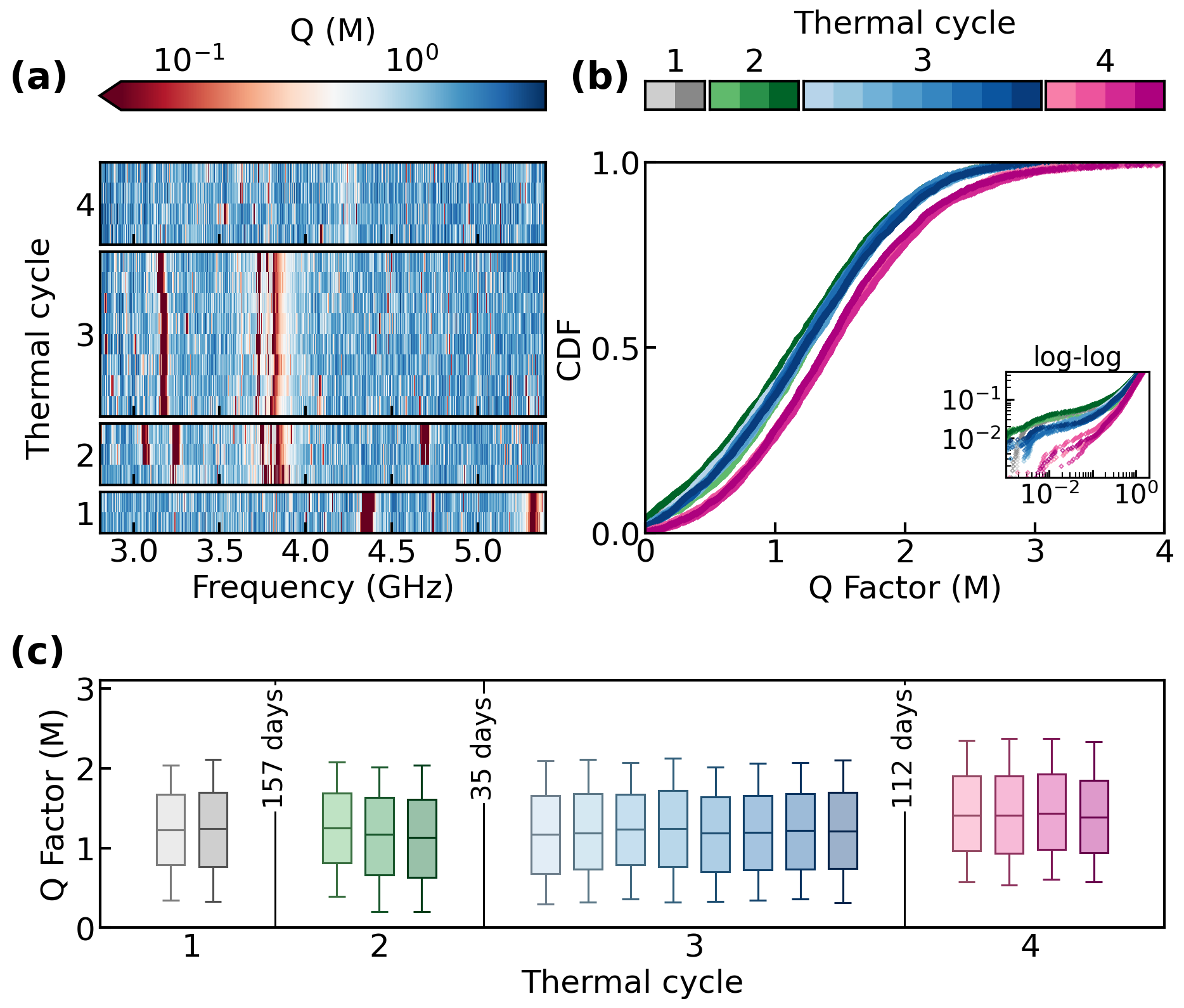}
\caption{\protect\input{figure3_caption.tex}}
\label{fig:fig3}
\end{figure}

While the broadband $Q$ distribution is repeatable within one thermal cycle despite TLS spectral dynamics, the few strongly coupled TLSs can drift outside the measurement bandwidth or fail to reappear after a thermal cycle (Fig.~\ref{fig:fig3}).
We collected 17 broadband sweeps across 326~days and four thermal cycles labeled TC\#1 to TC\#4, which add three cycles to the TC\#3 broadband data of Fig.~\ref{fig:fig2}.
Each boundary is a warm-up to room temperature except TC\#2 to TC\#3, an excursion to only 3.6~K.
TC\#4 was measured in a different dilution refrigerator.
The $Q$ spectra show the widest dropouts moving between thermal cycles, while the baseline stays constant in frequency and in time (Fig.~\ref{fig:fig3}a).
Accordingly, the broadband $Q$ distributions remain stable across TC\#1, TC\#2 and TC\#3, their CDFs overlapping tightly at both high and low $Q$ (Fig.~\ref{fig:fig3}b, main panel and log-log inset).
The widest dropouts are absent in TC\#4 (Fig.~\ref{fig:fig3}a), whose CDFs are offset at higher $Q$ than those of TC\#1 to TC\#3 (Fig.~\ref{fig:fig3}b).

We quantified how this reshuffling of the strongly coupled TLSs affects the broadband $Q$ CDFs and the medians drawn from them.
The medians spread by 10\% across TC\#1 to TC\#3, and TC\#4 sits 0.21~M above TC\#3, widening that spread to 24\%, from 1.13 to 1.43~M (Fig.~\ref{fig:fig3}c).
The median KS distance between the CDFs of TC\#3 and TC\#4 is 0.13, against 0.03 between repeats within either cycle.
Tightening the cut to $\chi^2_r < 2$ discards the fits in the low-$Q$ regions around the TLS resonances, reducing the spread to 12\%, from 1.33 to 1.50~M (see Sec.~\ref{sec:supp_redchi}).
The same cut reduces the median KS distance between TC\#3 and TC\#4 to 0.06 and leaves the distance between repeats at 0.03.
Computed in a sliding frequency window narrower than the full bandwidth, the KS distance localizes the disagreement between sweeps of TC\#3 and TC\#4 to the strongly coupled TLS resonances of TC\#3 (see Sec.~\ref{sec:supp_ks}).
Away from those resonances the CDFs overlap tightly, so a sufficiently broadband sweep separates the non-stationary contribution of the strongly coupled TLSs from the stationary baseline.

\begin{figure}[!t]
\centering
\includegraphics[width=0.95\columnwidth]{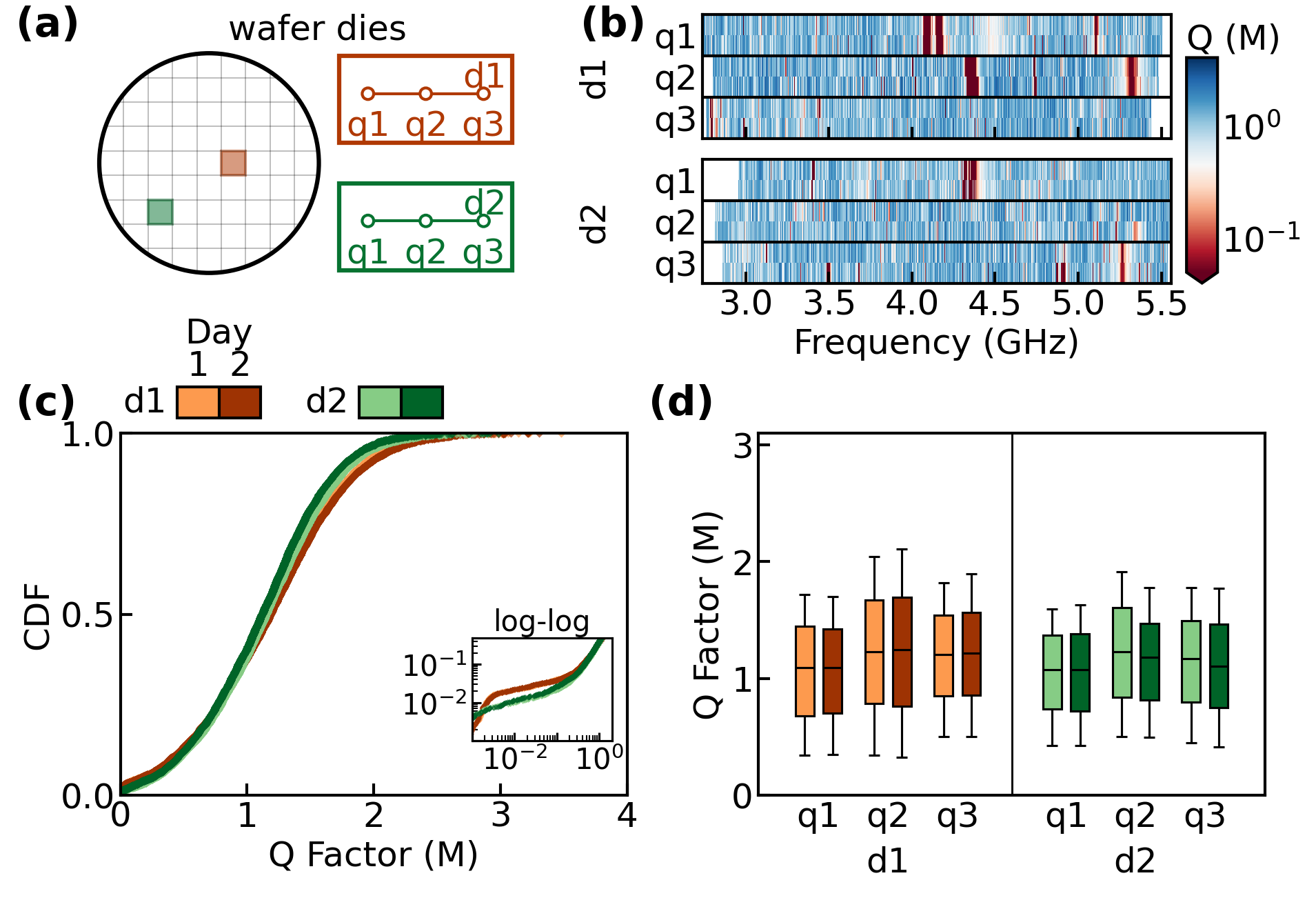}
\caption{\protect\input{figure4_caption.tex}}
\label{fig:fig4}
\end{figure}

We then used the broadband $Q$ distribution to compare qubit coherence across two dies of the same wafer (Fig.~\ref{fig:fig4}a), each hosting three identically designed qubits, all measured in the same thermal cycle.
Their six broadband $Q$ spectra share a common band of $\sim$2.5~GHz, up to fabrication variation in the maximum and minimum $f_{01}$ (Fig.~\ref{fig:fig4}b).
We repeated each sweep on the following day, which tests its repeatability and exposes the TLS spectral dynamics.
Each $Q$ CDF pools three qubits per die and day (Fig.~\ref{fig:fig4}c), and the four curves overlap tightly at high and low $Q$ (main panel and log-log inset), their medians spread by 0.06~M.
The KS distance between the two dies is 0.04 and 0.08 on the two days, against 0.01 and 0.04 for the 1-day repeats of d1 and d2.
Resolved by qubit and day (Fig.~\ref{fig:fig4}d), the 1-day repeat moves each median by at most 6\%, while the two dies differ by 10\% at a fixed qubit index, both below the 14\% spread across all twelve sweeps, whose medians run from 1.08 to 1.24~M.
Therefore, one-time sweeps on several copies of a qubit resolve fabrication uniformity above the variation of repeated sweeps on one copy.

In conclusion, we have established the $Q$ distribution pooled from a single broadband $Q$ spectrum as an energy-relaxation metric stationary within and across thermal cycles.
We applied the FFA SWAP spectroscopy sequence to probe energy relaxation with fine frequency resolution over a broad bandwidth.
Pooling the $Q$ spectrum into a CDF made stationarity testable through the KS distance between repeats, measured on equal-size samples at different bandwidths.
Only the broadband $Q$ CDFs came close to the sampling floor of the KS distance.
Across thermal cycles the pooled $Q$ medians varied by 10\% to 24\%, the larger spread following a reshuffling of the strongly coupled TLS resonances, and the broadband spectrum confirmed the $Q$ baseline unchanged away from them.
Between qubits on two dies of the same wafer, broadband $Q$ distributions resolved coherence differences that single-frequency or narrowband measurements would have masked.

Applied to materials development~\cite{Bal2024,Murthy2025,Biznarova2024,Joshi2026,Mahuli2025}, our method trades the simplicity of a fixed-frequency transmon for a flux-tunable one and its flux control line.
For qubits with longer $T_1$~\cite{Place2021,Wang2022,Bland2025}, each delay point will need a longer recycle time for $P_1$ to decay, lengthening a single sweep.
A device built for characterization could instead trade baseline $T_1$ for speed, narrowing the capacitor gaps to increase the coupling to the TLS bath and to the individual defects at the surface and metal interfaces, whose loss features then broaden enough for a coarser frequency step.
Smaller Josephson junctions would lower the chance that a strongly coupled TLS falls inside the swept band~\cite{Martinis2005}.
Packaging loss, Purcell decay into the readout resonator and resonances of the control lines will also set new limits on the $Q$ baseline~\cite{Martinis2014,Klimov2018}, appearing as static features of the $Q$ spectrum.
These trade-offs will guide the design of devices for broadband $Q$ characterization as qubit quality improves.

\begin{acknowledgments}
We thank the staff from across the Amazon Center for Quantum Computing who enabled this project.
A patent application related to this work has been filed.
\end{acknowledgments}

\bibliography{references}

\clearpage
\onecolumngrid
\begin{center}
  \large\bfseries Supplemental Material
\end{center}
\medskip
\twocolumngrid

\setcounter{secnumdepth}{3}
\setcounter{section}{0}
\setcounter{figure}{0}
\setcounter{equation}{0}
\setcounter{table}{0}
\renewcommand{\thesection}{S\arabic{section}}
\renewcommand{\thefigure}{S\arabic{figure}}
\renewcommand{\theequation}{S\arabic{equation}}
\renewcommand{\thetable}{S\arabic{table}}
\renewcommand{\theHfigure}{S\arabic{figure}}

\suppressfloats
\renewcommand{\topfraction}{0.75}

\makeatletter
\@booleantrue\raggedcolumn@sw
\makeatother

\section{The FFA SWAP spectroscopy pulse sequence}
\label{sec:supp_ffa}

\begin{figure}[tp]
  \centering
  \includegraphics[width=\columnwidth]{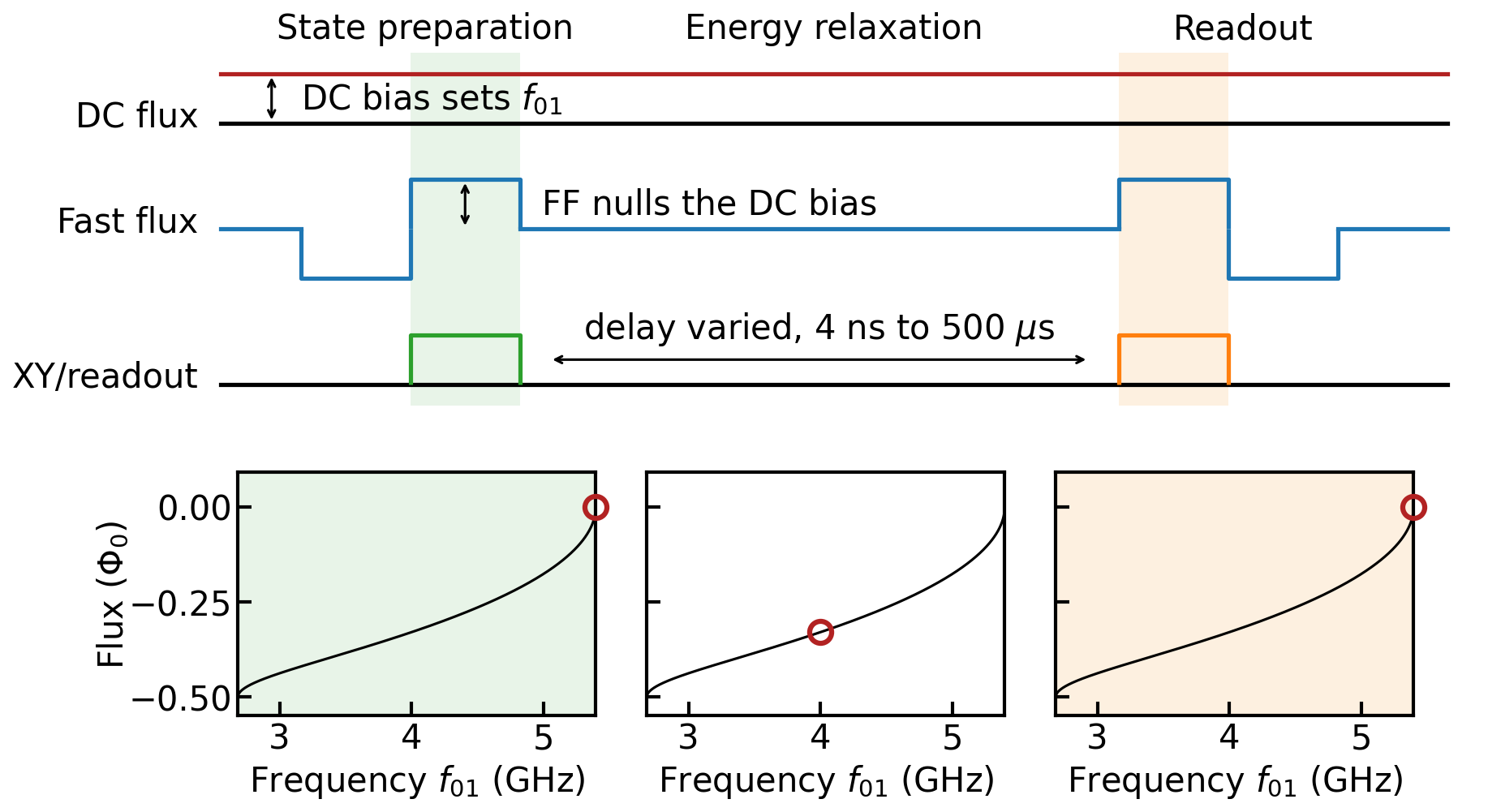}
  \caption{\protect\input{figureS1_caption.tex}}
  \label{fig:figS1}
\end{figure}

Probing the qubit energy-relaxation time $T_1$ over a broad bandwidth and with a fine frequency step requires stable control of the qubit frequency over times much longer than $T_1$ itself.
In conventional SWAP spectroscopy, a fast-flux (FF) pulse is held on while energy relaxation is probed.
However, state-of-the-art methods calibrate the distortion of such a pulse only on time scales up to tens of microseconds, leaving residual frequency errors of order 1~MHz~\cite{Hellings2025}.
To overcome this limitation and probe arbitrarily long $T_1$, we use the fast-flux-assisted (FFA) SWAP sequence of Ref.~\cite{Minguzzi2026}, shown in Fig.~\ref{fig:figS1}.
The DC bias sets and holds $f_{01}$ throughout, and two short FF pulses null that bias to bring the qubit back to the flux-insensitive point for state preparation and readout.
With the FF off during the delay, the qubit frequency is as stable as the DC source alone, and the FF pulse has to be calibrated only over time scales of $\leq$1~\textmu{}s.
The FF pulses are predistorted with IIR and FIR filters~\cite{Rol2020}, net-zero shaped and kept as short as possible so that residual transients do not introduce additional distortions.
Each $T_1$ curve is logarithmically sampled at 48 delays from 4~ns to 500~\textmu{}s.

\begin{figure}[tp]
  \centering
  \includegraphics[width=\columnwidth]{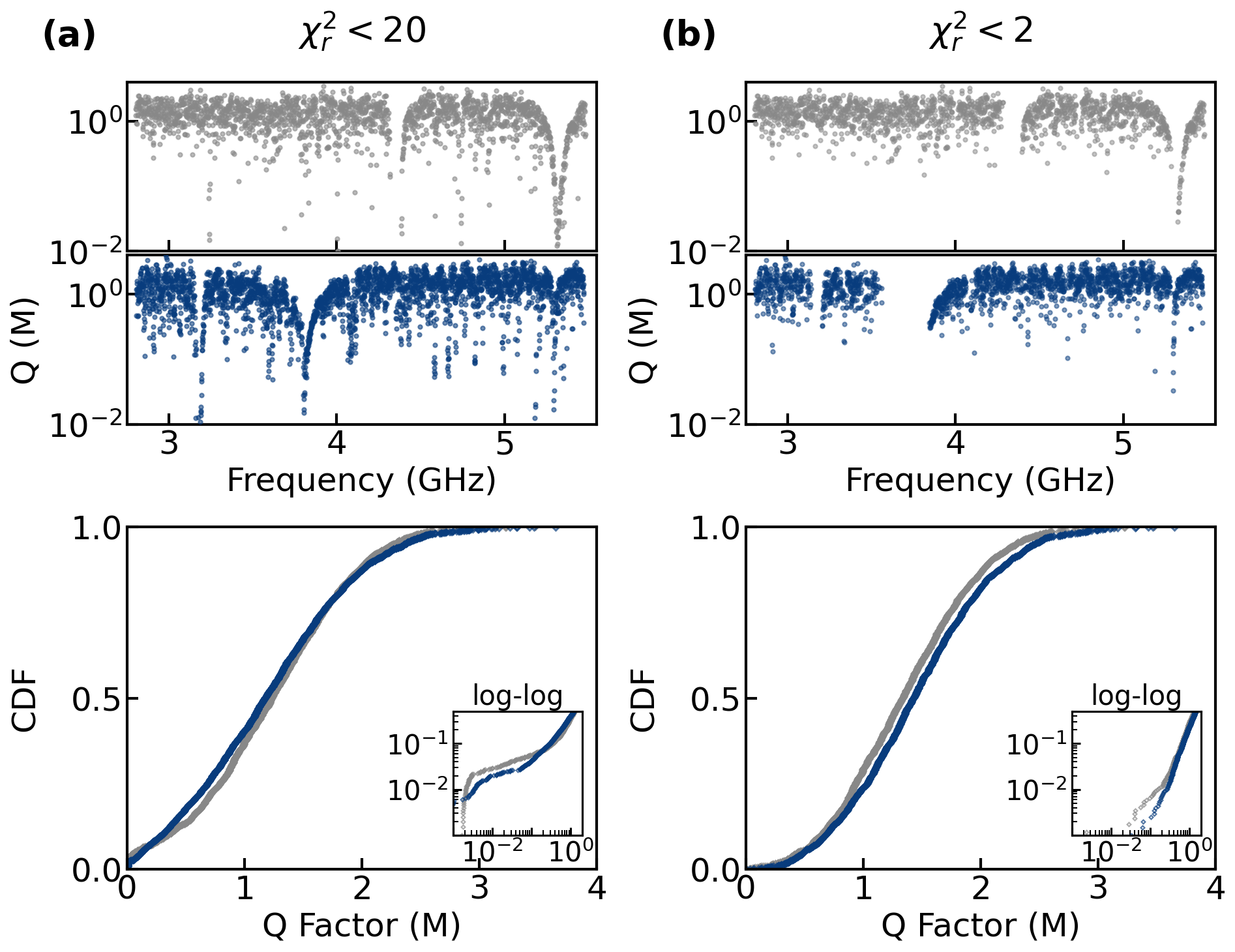}
  \caption{\protect\input{figureS2_caption.tex}}
  \label{fig:figS2}
\end{figure}

\begin{table}[tp]
  \squeezetable
  \caption{
    Fits discarded ($\Delta n$) and $Q$ median shift ($\Delta\mu$) from $\chi^2_r < 20$ to $\chi^2_r < 2$, averaged per cycle of Fig.~\ref{fig:fig3}.}
  \label{tab:supp_redchi_change}
  \begin{ruledtabular}
  \begin{tabular}{lccc}
    Cycle & Sweeps & $\Delta n$ (\%) & $\Delta\mu$ ($10^6$) \\
    \hline
    TC\#1 & 2 & 14 & +0.11 \\
    TC\#2 & 3 & 30 & +0.24 \\
    TC\#3 & 8 & 30 & +0.23 \\
    TC\#4 & 4 & 8 & +0.07 \\
  \end{tabular}
  \end{ruledtabular}
\end{table}

\begin{table}[t]
  \squeezetable
  \caption{
    Effect of the two fit quality cuts on the stationarity of the sweeps across the thermal cycles of Fig.~\ref{fig:fig3}.
    The median spread rows show the minimum and maximum per-sweep median, with relative percentage change in parentheses.
    Each KS distance $D$ is a median over pairs of sweeps.}
  \label{tab:supp_redchi_spread}
  \begin{ruledtabular}
  \begin{tabular}{lcc}
     & $\chi^2_r < 20$ & $\chi^2_r < 2$ \\
    \hline
    \multicolumn{3}{l}{Median spread ($10^6$)} \\
    ~~TC\#1--TC\#4         & 1.13--1.43 (24\%) & 1.33--1.50 (12\%) \\
    ~~TC\#1--TC\#3         & 1.13--1.25 (10\%) & 1.33--1.45 (8\%) \\
    \hline
    \multicolumn{3}{l}{KS distance $D$} \\
    ~~within TC\#3         & 0.03 & 0.03 \\
    ~~within TC\#4         & 0.03 & 0.03 \\
    ~~TC\#3 vs TC\#4       & 0.13 & 0.06 \\
  \end{tabular}
  \end{ruledtabular}
\end{table}

\begin{figure}[t]
  \centering
  \includegraphics[width=\columnwidth]{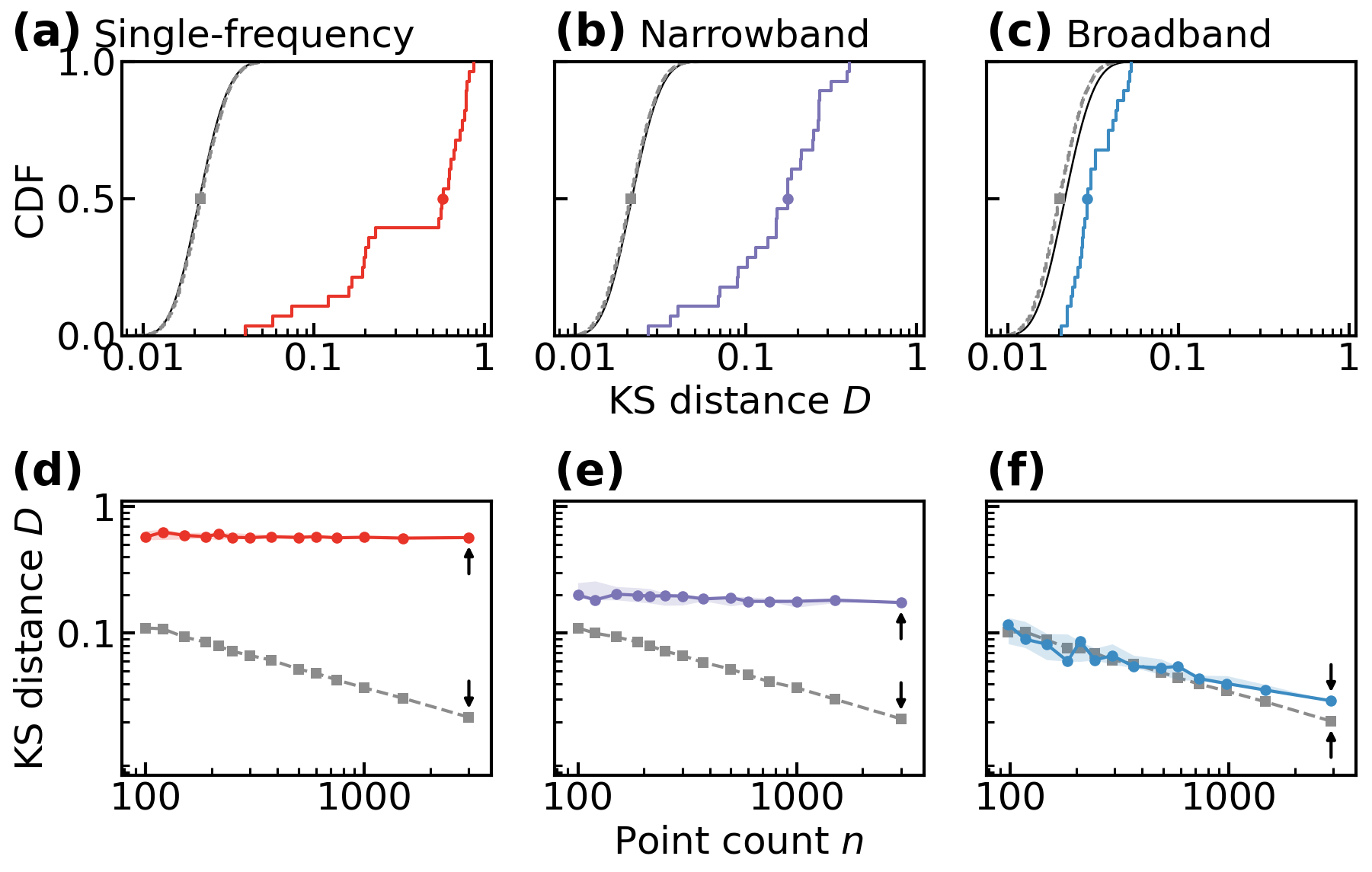}
  \caption{\protect\input{figureS3_caption.tex}}
  \label{fig:figS3}
\end{figure}

\begin{figure}[t]
  \centering
  \includegraphics[width=\columnwidth]{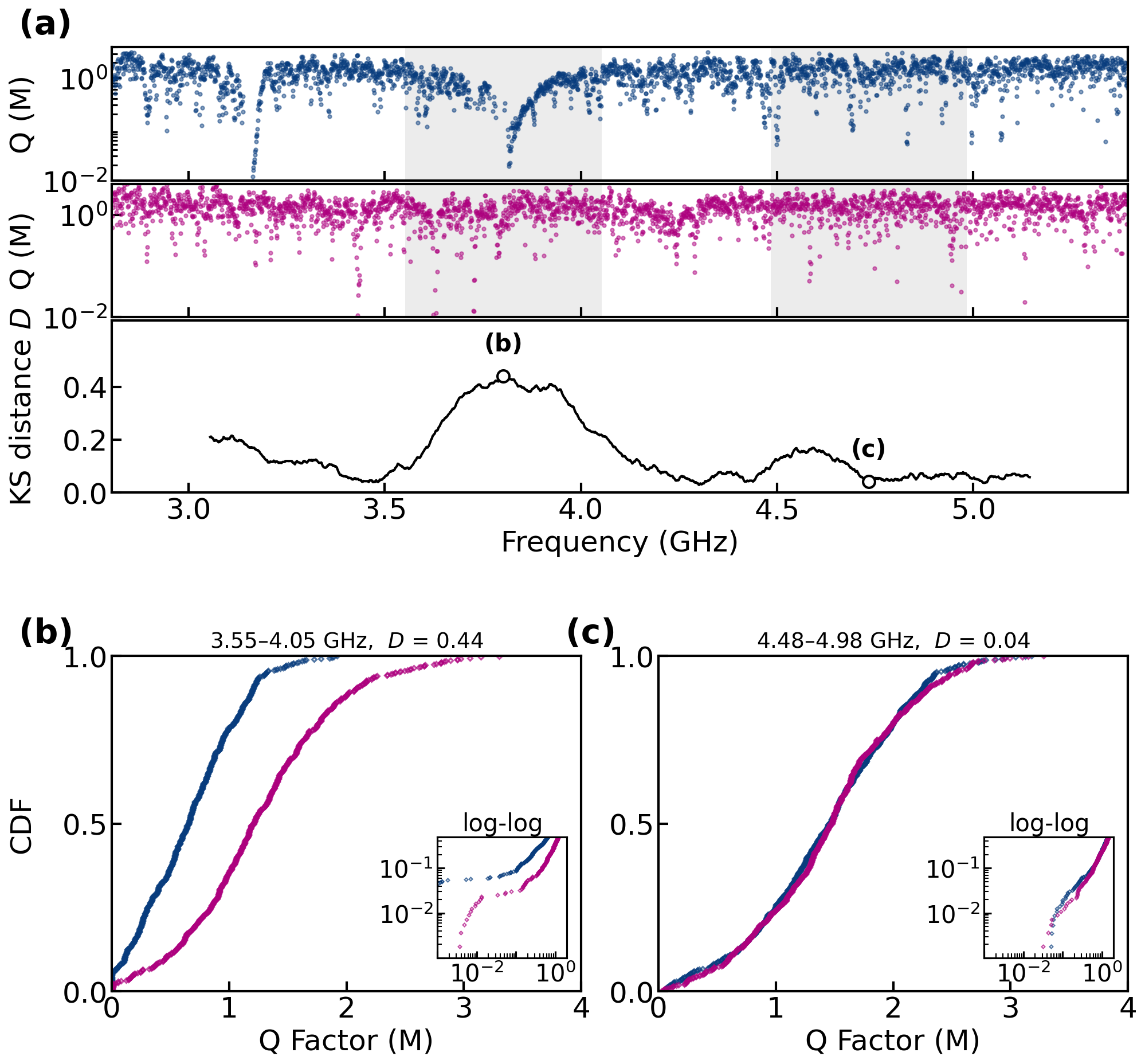}
  \caption{\protect\input{figureS4_caption.tex}}
  \label{fig:figS4}
\end{figure}

\section{Sensitivity of the \texorpdfstring{$Q$}{Q} distribution to the fit quality filters}
\label{sec:supp_redchi}

Each quoted $Q = 2\pi f_{01} T_1$ comes from the $T_1$ of an exponential decay fit, whose reduced chi-squared $\chi^2_r$ can be used as a quality filter.
For example, a large $\chi^2_r$ can arise from non-exponential decay when the qubit exchanges energy coherently with a strongly coupled TLS.
In Fig.~\ref{fig:figS2} we illustrate how different cuts on $\chi^2_r$ affect the $Q$ spectra and CDFs of thermal cycles TC\#1 and TC\#3.
For both sweeps, the stricter cut $\chi^2_r < 2$ removes most of the points with $Q < 0.1$~M, which is visible as blank regions around the largest TLS resonances in the spectra (Fig.~\ref{fig:figS2}, top rows) and as the shorter low-$Q$ tail of the CDFs (log-log insets).
The fraction of points removed is about 14\% and 30\%, and the medians rise by 0.11 and 0.23~M, respectively (Table~\ref{tab:supp_redchi_change}).
The unequal rise of the medians is sufficient to change their relative difference, so that TC\#3 has the higher median when $\chi^2_r < 2$ (Fig.~\ref{fig:figS2}, bottom rows).
A stricter cut can therefore change the relative median spread between two distributions, which reflects the low-$Q$ points it has removed.

We then apply the same stricter cut to all 17 broadband sweeps of Fig.~\ref{fig:fig3}, which test stationarity across four thermal cycles and the effect of the strongly coupled TLSs reshuffling between the cycles (Tables~\ref{tab:supp_redchi_change} and~\ref{tab:supp_redchi_spread}).
Under $\chi^2_r < 20$, the absence of the widest TLS resonances in TC\#4 widens the median spread from 10\%, across TC\#1 to TC\#3, to 24\% across all four cycles.
The fraction of fits discarded follows the strongly coupled TLS population of each cycle, 30\% in TC\#3 against 8\% in TC\#4, and the medians rise by 0.23 and 0.07~M, respectively.
The two spreads therefore narrow, to 8\% across TC\#1 to TC\#3 and 12\% across all four cycles, and the median offset between TC\#3 and TC\#4 falls from 0.21~M to 0.06~M.
The median KS distance between the two cycles falls from 0.13 to 0.06, while the distance within either cycle stays at 0.03.
The stationarity of Fig.~\ref{fig:fig3} is therefore not an artifact of the fit quality filter, and we keep $\chi^2_r < 20$ throughout the main text.

\section{Quantifying repeatability with the Kolmogorov-Smirnov distance}
\label{sec:supp_bandwidth_ks}

We used the two-sample Kolmogorov-Smirnov (KS) distance $D$ to test the repeatability of sweeps at different bandwidths.
The eight repeats at each bandwidth in Fig.~\ref{fig:fig2} give 28 pairs, whose distances we plot as cumulative distributions in Figs.~\ref{fig:figS3}a-c, one panel per bandwidth.
Each repeat is a 3{,}000-point sweep, of which we keep the points that pass the fit quality filters of Sec.~\ref{sec:supp_redchi} in all eight sweeps.
The median $D$ falls from 0.57 at a single frequency to 0.17 over the 50~MHz band (narrowband) and to 0.03 over the 2.7~GHz band (broadband), the circles of Figs.~\ref{fig:figS3}a-c.
For two independent draws of point counts $n,m$, the distribution of $D$ would follow the Kolmogorov law (black curves) with a median of 0.83\,$\sigma_0$ and $\sigma_0 = \sqrt{1/n + 1/m}$.
Rather than assume independent draws, we measured the sampling floor of each bandwidth by permuting the repeat labels independently at each point of the shared sampling grid and recomputing all 28 distances, over 300 replicates~\cite{Feigelson2012}.
The median of each resulting distribution (dashed grey curves) is the sampling floor of that bandwidth (grey squares).
For the single-frequency and narrowband cases this floor lies close to the Kolmogorov law, while the broadband floor falls below it at 0.77\,$\sigma_0$, which indicates that $Q$ varies more with frequency than between repeats.
The median $D$ is therefore 26, 8.3 and 1.4 times its own sampling floor, and only the broadband sweeps come close to it.

We repeated this comparison at lower point counts by subsampling the same grid of each sweep (Figs.~\ref{fig:figS3}d-f).
Taking every $k$th point of the grid gives $k$ equally plausible sweeps of $n/k$ points over the same band.
For each bandwidth, we recomputed the sampling floors at 14 values of $k$, from close to 3{,}000 points down to about 100.
The median $D$ is flat at $\sim$0.6 at a single frequency and $\sim$0.2 over the 50~MHz band at all point counts, while over the 2.7~GHz band it rises from 0.03 to 0.12, tracking its own sampling floor, which rises as $1/\sqrt{n}$.
Circles show one of the $k$ sweeps at each density and the shaded bands span them all, reaching the sampling floor only over the 2.7~GHz band (Fig.~\ref{fig:figS3}f).
This is not closer agreement between repeats at a lower point count, but a measurement whose sampling floor has grown past the difference it can resolve.

\section{Distinguishing a local TLS from a global change of the \texorpdfstring{$Q$}{Q} distribution}
\label{sec:supp_ks}

The KS distance $D$ evaluated over the entire bandwidth cannot distinguish an overall change of the $Q$ distribution from a local change due to a strongly coupled TLS.
We separate the two effects by evaluating $D$ in a common frequency window stepped across the band.
We illustrate this in Fig.~\ref{fig:figS4}a with one broadband sweep from TC\#3 (blue shade) and another from TC\#4 (pink shade), two cycles whose $Q$ CDFs are separated by a clear gap in Fig.~\ref{fig:fig3}b.
We compute $D$ over a moving 500~MHz window across the full band, where it varies from 0.03 to 0.44 (Fig.~\ref{fig:figS4}a, bottom).
The maximum, at 3.80~GHz, is the window centered on a strongly coupled TLS present in TC\#3 but absent in TC\#4, and we select for comparison a window of small $D$ far from it, 4.48 to 4.98~GHz (shaded bands in Fig.~\ref{fig:figS4}a).
The $Q$ CDFs show weak (Fig.~\ref{fig:figS4}b) and tight overlap (Fig.~\ref{fig:figS4}c), respectively.
Normalizing $D$ by the sampling scale $\sigma_0$ of Sec.~\ref{sec:supp_bandwidth_ks}, the $D/\sigma_0 = 0.7$ in that window is much smaller than the $D/\sigma_0 = 4.1$ over the full band shared by the two sweeps, where $D = 0.11$.
The disagreement is therefore confined to the neighborhood of the strongly coupled TLS, and away from it the $Q$ baseline is unchanged across the two thermal cycles.

\end{document}